# Cooperative photoinduced metastable phase control in strained manganite films

J. Zhang[1,3*], X. Tan[2], M. Liu[3,5], S. W. Teitelbaum[4], K.W. Post[3], Feng Jin[2], K. A. Nelson[4], D. N. Basov[3], W. Wu[2], R.D. Averitt[1,3*]

[1] Department of Physics, Boston University, Boston, MA 02215, USA.

[2] Hefei National Laboratory for Physical Sciences at Microscale, and High Magnetic Field Laboratory of Chinese Academy of Sciences, University of Science and Technology of China, Hefei, Anhui 230026, China.

[3] Department of Physics, University of California at San Diego, La Jolla, CA 92093, USA.

[4] Department of Chemistry, Massachusetts Institute of Technology, Cambridge MA 02139, USA.

[5] Department of Physics, Stony Brook University, Stony Brook, New York 11790, USA.

[*]To whom correspondence should be addressed.
E-mail: jdzhang@physics.ucsd.edu (J.Z.); raveritt@ucsd.edu (R.D.A.)

**A major challenge in condensed matter physics is active control of quantum phases. Dynamic control with pulsed electromagnetic fields can overcome energetic barriers enabling access to transient or metastable states that are not thermally accessible. Here we demonstrate strain-engineered tuning of $La_{2/3}Ca_{1/3}MnO_3$ into an emergent charge-ordered insulating phase with extreme photo-susceptibility where even a single optical pulse can initiate a transition to a long-lived metastable**

**hidden metallic phase. Comprehensive single-shot pulsed excitation measurements demonstrate that the transition is cooperative and ultrafast, requiring a critical absorbed photon density to activate local charge excitations that mediate magnetic-lattice coupling that, in turn, stabilize the metallic phase. These results reveal that strain engineering can tune emergent functionality towards proximal macroscopic states to enable dynamic ultrafast optical phase switching and control.**

Precision tuning of the local environment in complex materials provides a route to control macroscopic functionality thereby offering a glimpse into how microscopic interactions conspire toward emergent behavior. Tuning can be achieved through doping, heterostructuring, epitaxial strain, or with applied static or dynamic electromagnetic fields, which includes photoexcitation [1,2]. Examples include the melting of stripe order in a high-$T_c$ cuprate in favor of superconductivity, the generation of a hidden phase in a layered dichalcogenide, and the observation of Floquet states in a topological insulator [3,4,5]. Transition metal oxides (TMOs) are of considerable interest with regards to dynamic tuning because of structure-function coupling arising, predominantly, from octahedral rotations and distortions [6,7,8]. The rich structure-function coupling results in a complex energy landscape with distinct optical signatures that facilitate identification of dynamic phase changes [9,10]. Doped rare-earth perovskite manganites, derived from the Mott-insulating parent compound $ReMnO_3$ (Re = Pr, La, Nd), have proven fertile in the search for novel dynamics with a focus on the insulator-to-metal transition [11,12,13,14,15] that is strongly influenced from coupling between the lattice and the underlying electronic and magnetic structure[16,17,18,19].

A state-of-the-art technique to mediate transition metal oxide properties is elastic strain engineering through epitaxial growth, from which new electronic states can emerge [20,21]. We effectively hide the low-temperature ferromagnetic metallic (FM) phase present in unstrained $La_{2/3}Ca_{1/3}MnO_3$ (LCMO, $T_c \approx 260K$) in favor of an antiferromagnetic insulating (AFI) phase[22,23] utilizing the anisotropic strain from $NdGaO_3$ (001) substrates (**see Supplementary information**). The AFI phase originates from enhanced orthorhombicity of the LCMO lattice (**Fig. 1b**) where octahedral tilting and strain-enhanced Jahn-Teller-like distortion breaks the cubic symmetry, decreasing the Mn-O-Mn bond angle. This suppresses the ferromagnetism and d-electron itinerancy resulting in a low-temperature charge-ordered insulating state that evolves toward the conventional paramagnetic insulating phase at higher temperatures.

**Figure 1a** shows resistivity measurements for a strained 30 nm LCMO film. In zero applied magnetic field, the film remains insulating at all temperatures (due to strain-enhanced orthorombicity[22,23] as depicted in **Fig. 1d**). For fields above 3 T the insulating phase collapses becoming a ferromagnetic metal at low temperatures. **Figure 1b** details the phase diagram of strained LCMO as determined from the field-dependent transport measurements [22,23]. The FM and AFM phases coexist over the range from 0-3 T, depending on temperature. To characterize the time-integrated electrodynamic response of the strained films, the optical conductivity was measured from 100 meV to 5 eV using spectroscopic ellipsometry (**Fig. 1c**). With decreasing temperature, the film displays spectral weight transfer from a small polaron (~1.5 eV) peak to a sharp well-defined peak (1.7 eV) suggestive of charge order and associated with intersite transitions $Mn^{3+}$-$Mn^{4+}$ ($d_i^3 d_j^4 \rightarrow d_i^4 d_j^3$) as depicted in **Fig. 1e** [24][25][26][27]. This is in stark contrast to the development of a coherent Drude response with decreasing temperature in unstrained films (**Supplementary Fig. 3a**).

As we now show, photoexcitation recovers the hidden FM phase of the strain-engineered insulating LCMO films. The films were excited with 1.55 eV photon energy pulses (35 fs, see **Supplementary information** for details) resonantly exciting intersite $Mn^{3+}$-$Mn^{4+}$ transitions (**Fig. 1d, e**). Terahertz (THz) time domain spectroscopy was used to measure the photoinduced change in conductivity. The photoinduced phase transition from the insulating state to the "hidden" ferromagnetic metallic state is shown in **Figure 2** and constitutes the main result of this work. In the absence of photoexcitation, there is no measureable THz conductivity, consistent with an insulating state and the optical conductivity measurements in **Fig. 1c**. In **Figure 2**, the temperature of the LCMO is decreased from 120 K to 80 K in 5 K decrements. Two measurements of THz conductivity, before and after 1.55 eV photoexcitation (fluence ~ 2 mJ/cm$^2$), were performed at each temperature of the cooling process. At temperatures lower than 105 K, a persistent change of THz conductivity occurs after photoexcitation. When the sample was cooled by an additional 5 K (in the absence of photoexcitation), the conductivity remained constant, and only increased upon additional photoexcitation. The conductivity continued to increase and formed a step-like curve as a function of decreasing temperature until it reached the full metallic state conductivity of ~1,200 Ω$^{-1}$cm$^{-1}$ at 80 K. Subsequently, the conductivity was measured with increasing temperature in the absence of photoexcitation. A monotonic decrease was observed with a return to the insulating state at 130 K. After finishing the thermal cycle, the THz conductivity was measured without any optical excitation. The LCMO film again remained insulating at all temperatures. The entire process could be repeated without permanent changes in the optical properties of the film.

The photoinduced THz conductivity (PTC) is stable as long as the temperature is maintained. Further, the maximum conductivity is the same value as obtained with a strong magnetic field. This is clear from the blue dots of **Fig. 1a**, where the PTC from **Fig. 2** is mapped onto the temperature-dependent DC resistivity curve. That is, optical excitation results (at 80 K) in a metallic state conductivity equivalent to that accessed with magnetic fields higher than 3 T. The correspondence of the metallic state, obtained through two different pathways, is indicative of dynamic coupling between the charge, spin, and lattice degrees of freedom (DOF). The observed laser-induced IMT is a nonequilibrium dynamical process and not simply laser heating, since the metallic phase is not thermally accessible. In fact, laser heating drives the film back toward the insulating phase.

To further investigate the nature of the photoinduced metastable metallic state, single-shot experiments were performed where, on a shot by shot basis, the conductivity change was measured (details in **Supplementary information**). **Figure 3a** plots the resistivity (as determined from the THz probe) as a function of the number of absorbed pulses at 80 K with a fluence of 4 mJ/cm$^2$. Following the absorption of the first pulse, the resistivity decreases by approximately three orders of magnitude. With subsequent pulses, the resistivity continues to decrease by an additional order of magnitude, reaching a minimum after ~20 pulses. **Figure 3b** plots the photoinduced conductivity versus shot number for different fluences. The 4 mJ/cm$^2$ (same data as **Fig. 3a**) saturates at 800 $\Omega^{-1}$cm$^{-1}$, while at lower fluences the conductivity saturates at a lower value. This is important, showing that the conductivity change does not simply arise from the absorbed number of photons. If there were a simple dependence on the number of absorbed photons, the data at lower fluences would saturate to the same conductivity value as the higher fluence data after a sufficient number of pulses. This is clearly not the case and indicates a cooperative process with a photon-absorption threshold. **Figure 3c** plots the conductivity plateaus from **Fig. 3b** versus the number of absorbed photons (as determined from the fluence and optical conductivity data), yielding a threshold of approximately 0.06 photons per active Mn$^{3+}$ to stabilize the metallic phase.

The photoinduced conductivity can take on a range of values indicating FM and AFI phase coexistence, where photon absorption results in spectral weight transfer from the visible to the far-infrared corresponding to an increase of the FM volume fraction at the expense of the AFI volume fraction. Using effective medium theory (see **Supplementary information**) we have determined for different fluences the rate of increase (per laser pulse absorbed) of the FM volume fraction. From this analysis, exponential growth (see equation (6), **Supplementary information**) of the FM conductivity is expected with the resultant fits shown (solid lines) in **Figure 3b**. Starting from the AFI state, the first absorbed pulse drives the LCMO films beyond the percolation threshold, with subsequent pulses further increasing the metallic volume fraction.

Measuring the conductivity dynamics of the pristine AFI state following single-pulse excitation would provide insight into the photoinduced IMT but is not experimentally feasible on a shot-to-shot basis. Instead, we employed an all-optical single-shot ultrafast spectroscopic method[28] which faithfully represents the conductivity dynamics because of the aforementioned spectral weight transfer. **Figure 3d** shows the results of single-shot photoinduced reflectivity dynamics ($\Delta R/R$) probing at the peak of the intersite transition (1.7 eV) following 1.55 eV excitation. There is a decrease in $\Delta R/R$ consistent with dynamic spectral weight transfer to THz frequencies and the IMT dynamics occur on a sub-picosecond timescale with no evidence of longer time-scale dynamics. This is consistent with an ultrafast nonthermal cooperative process driving the IMT. We have also monitored the spatial evolution of photo-induced metallic phase on a shot-to-shot basis (**Fig. 3 e-j**), by taking optical images of the photoexcited area with the laser pulse focused to 350 μm (FWHM) in diameter indicated by the dashed circle. The optical reflectivity of the switched region shows a spatial dependence, due to the Gaussian beam intensity profile with the highest fluence at its center. Interestingly, even at the excitation center, the images clearly show the coexistence of micron scale regions with different reflectivity, consistent with phase coexistence.

We now consider how photoexcitation initiates the dynamics that result in a stabilized metallic phase. Photon absorption at 1.55 eV delocalizes the $e_g$ d-electrons of the $Mn^{3+}$ atoms (**Fig. 1d, e, Supplementary Fig. 8**) and drives, because of strong electron-phonon coupling, preferential relaxation of the Jahn-Teller distortions of the $MnO_6$ octahedra. This can happen within a fraction of a phonon period, consistent with the ultrafast dynamics observed in the single-shot measurements. In turn, the rapid lattice reorientation directly affects the value of the exchange integral leading to a transition from antiferromagnetic to ferromagnetism which can, in principle, preserve the itinerancy of d-electrons in the $e_g$ band and the relaxed lattice structure[29], due to the strong magnetic-lattice coupling. However, at low fluences, only a small portion of the $e_g$ d-electrons delocalize, and the corresponding density of relaxed octahedra is low, hindering the onset of a global phase transition. At higher pump fluences, the population of transiently relaxed octahedra increases such that above ~0.06 photons per active $Mn^{3+}$ the metallic state is favored. We note that the PTC shows a strong dependence on the excitation pulse duration exhibiting a threshold of 200 fs (e.g. 2 $mJ/cm^2$ pulses longer than 200 fs cannot switch the sample). This further indicates that the PTC is mediated by coherence effects (i.e. displacive excitation of optical phonons).

Phenomenological Ginzberg-Landau (GL) modeling (see **Supplementary informaition**) suggests that magnetic-lattice (ML) coupling[30] is a crucial ingredient to obtain a metastable FM phase that persists over a well-defined range of temperatures in the absence of any sustained external influence. This is shown in **Figure 4**, which plots the free energy for the case of weak and strong coupling between the lattice and the magnetization order parameters. In the limit of weak ML coupling **(Fig. 4a, c)**, distinct minima can exist for the lattice distorted state (A) and the ferromagnetically ordered state (B). A photoexcited trajectory going from A to B as described above is feasible. However the shallow minimum of the metallic ferromagnetic state would, from thermal fluctuations, result in a rapid return to the lattice-distorted insulating phase. With increasing ML coupling, as depicted in **Fig. 4 b, d**, the magnetic order has a deeper local minimum protecting it from thermal fluctuations over an increased temperature range. This suggests that strong ML coupling plays a crucial role in establishing a robust, yet metastable, ferromagnetic metallic state. Nonetheless, biquadratic ML coupling is insufficient to explain all of the features of our data. For example, phase separation may arise from bilinear coupling between orthorhombic and Jahn-Teller distortions, i.e. acoustic and optic phonon coordinates[17]. A more complete description of the photoinduced IMT would need to include this in addition to ML coupling.

Our results suggest that strain-engineering of transition metal oxides to precisely control the atomic lattice provides a powerful approach to encode new emergent electronic phases that are robust yet extremely sensitive to dynamic perturbation with optical excitation. This opens a new rou te to design functional optoelectric devices where the tunable macroscopic properties derive from strongly interacting degrees of freedom present in transition metal oxides.

**References:**

1.  Nasu, K. *Photoinduced Phase Transitions*. (World Scientific, 2004).

2. Zhang, J. & Averitt, R. D. Dynamics and Control in Complex Transition Metal Oxides. *Ann. Rev. Mat. Res.* **44,** 19-43 (2014).

3. Fausti, D. *et al.* Light-induced superconductivity in a stripe-ordered cuprate. *Science* **331,** 189–91 (2011).

4. Stojchevska, L. *et al.* Ultrafast Switching to a Stable Hidden Quantum State in an Electronic Crystal. *Science* **344,** 177–180 (2014).

5. Wang, Y. H., Steinberg, H., Jarillo-Herrero, P. & Gedik, N. Observation of Floquet-Bloch states on the surface of a topological insulator. *Science* **342,** 453–457 (2013).

6. Rondinelli, J. M., May, S. J. & Freeland, J. W. Control of octahedral connectivity in perovskite oxide heterostructures: An emerging route to multifunctional materials discovery. *MRS Bulletin* **37,** 261–270 (2012).

7. Imada, M., Fujimori, A. & Tokura, Y. Metal-insulator transitions. *Rev. Mod. Phys.* **70,** 1039-1263 (1998).

8. May, S. J. *et al.* Quantifying octahedral rotations in strained perovskite oxide films. *Phys. Rev. B* **82,** 14110 (2010).

9. Liu, M. *et al.* Terahertz-field-induced insulator-to-metal transition in vanadium dioxide metamaterial. *Nature* **487,** 345–8 (2012).

10. Rini, M. *et al.* Control of the electronic phase of a manganite by mode-selective vibrational excitation. *Nature* **449,** 72–4 (2007).

11. Miyano, K., Tanaka, T., Tomioka, Y. & Tokura, Y. Photoinduced Insulator-to-Metal Transition in a Perovskite Manganite. *Phys. Rev. Lett.* **78,** 4257–4260 (1997).

12. Takubo, N. *et al.* Persistent and Reversible All-Optical Phase Control in a Manganite Thin Film. *Phys. Rev. Lett.* **95,** 017404 (2005).

13. Ichikawa, H. *et al.* Transient photoinduced 'hidden' phase in a manganite. *Nature Materials* **10,** 101–5 (2011).

14. Asamitsu, A., Tomioka, Y., Kuwahara, H. & Tokura, Y. Current switching of resistive states in magnetoresistive manganites. *Nature* **388,** 50–52 (1997).

15. Li, T. *et al.* Femtosecond switching of magnetism via strongly correlated spin-charge quantum excitations. *Nature* **496,** 69–73 (2013).

16. Hwang, H. Y., Cheong, S.-W., Radaelli, P. G., Marezio, M. & Batlogg, B. Lattice Effects on the Magnetoresistance in Doped LaMnO$_3$. *Phys. Rev. Lett.* **75,** 914 (1995).

17. Ahn, K., Lookman, T. & Bishop, A. Strain-induced metal–insulator phase coexistence in perovskite manganites. *Nature* **804,** 401–404 (2004).

18. Burgy, J., Moreo, A. & Dagotto, E. Relevance of Cooperative Lattice Effects and Stress Fields in Phase-Separation Theories for CMR Manganites. *Phys. Rev. Lett.* **92,** 097202 (2004).

19. Kiryukhin, V. *et al.* An X-ray-induced insulator--metal transition in a magnetoresistive manganite. *Nature* **386,** 813–815 (1997).

20. Chakhalian, J., Millis, a J. & Rondinelli, J. Whither the oxide interface. *Nature Materials* **11,** 92–94 (2012).

21. Nakamura, M., Ogimoto, Y., Tamaru, H., Izumi, M. & Miyano, K. Phase control through anisotropic strain in Nd$_{0.5}$Sr$_{0.5}$MnO$_3$ thin films. *Appl. Phys. Lett.* **86,** 182504 (2005).

22. Huang, Z. *et al.* Tuning the ground state of La$_{0.67}$Ca$_{0.33}$MnO$_3$ films via coherent growth on orthorhombic NdGaO$_3$ substrates with different orientations. *Phys. Rev. B* **86,** 014410 (2012).

23. Wang, L. F. *et al.* Annealing assisted substrate coherency and high-temperature antiferromagnetic insulating transition in epitaxial La$_{0.67}$Ca$_{0.33}$MnO$_3$/NdGaO$_3$ (001) films. *AIP Advances* **3,** 52106 (2013).

24. Basov, D. N., Averitt, R. D., van der Marel, D., Dressel, M. & Haule, K. Electrodynamics of correlated electron materials. *Rev. Mod. Phys.* **83,** 471–541 (2011).

25. Kovaleva, N. *et al.* Spin-Controlled Mott-Hubbard Bands in LaMnO$_3$ Probed by Optical Ellipsometry. *Phys. Rev. Lett.* **93,** 147204 (2004).

26. Quijada, M., Černe, J., Simpson, J. & Drew, H. Optical conductivity of manganites: Crossover from Jahn-Teller small polaron to coherent transport in the ferromagnetic state. *Phys. Rev. B* **58,** 16093–16102 (1998).

27. Kim, K., Lee, S., Noh, T. & Cheong, S.-W. Charge Ordering Fluctuation and Optical Pseudogap in La$_{1-x}$Ca$_x$MnO$_3$. *Phys. Rev. Lett.* **88,** 167204 (2002).

28. Shin, T., Wolfson, J. W., Teitelbaum, S. W., Kandyla, M. & Nelson, K. A. Dual echelon femtosecond single-shot spectroscopy. *Rev. Sci. Inst.* **85,** 83115 (2014).

29. Wall, S., Prabhakaran, D., Boothroyd, a. & Cavalleri, a. Ultrafast Coupling between Light, Coherent Lattice Vibrations, and the Magnetic Structure of Semicovalent LaMnO$_3$. *Phys. Rev. Lett.* **103,** 097402 (2009).

30. Chatterji, T., Henry, P. & Ouladdiaf, B. Neutron diffraction investigation of the magneto-elastic effect in LaMnO$_3$. *Phys. Rev. B* **77,** 212403 (2008).

**Acknowledgments:** RDA and JZ acknowledge support from DOE - Basic Energy Sciences under Grant No. DE-FG02-09ER46643. XT and WW acknowledge support from the NSF of China (Grant No. 11274287) and the National Basic Research Program of China (Grant Nos. 2012CB927402 and 2015CB921201). SWT and KAN acknowledge support from Office of Naval Research (N00014-12-1-0530) and the National Science Foundation (CHE-1111557).

**Author Contributions:** J.Z. and R.D.A. developed the idea. X.T., F.J. and W.W. performed material growth and film characterization. J.Z., M. L., K.W.P., D.N.B. performed the optical conductivity measurements. J.Z., and S.W.T., R.D.A., K.A.N. performed the THz and single shot measurements. J.Z. and R.D.A. performed the GL analysis. R.D.A, W.W., D.N.B., and K.A.N. supervised the project. J.Z. and R.D.A. wrote the paper. All authors contributed to the understanding of the physics and revised the paper.

**Competing financial interests**
The authors declare no competing financial interests.

**Corresponding authors**
Correspondence to: J. D. Zhang or R. D. Averitt.

**Figure 1 | Strain engineered La$_{2/3}$Ca$_{1/3}$MnO$_3$ (LCMO) thin film. a,** DC Resistivity-Temperature curves for 30nm strain-engineered LCMO film, under various magnetic fields (0~5 T). The blue dots are the conductivity following excitation with sub-50 fs 1.5 eV pulses under zero field (see text, and Fig. 2). **b,** Phase diagram of sample film where boundaries (AFI – antiferromagnetic insulator, FM – ferromagnetic metal) are extrapolated from magnetization and resistivity measurements (yellow dots). **c,** Optical conductivity of the film at various temperatures showing resonance peak at ~1.7 eV, corresponding to inter-site hopping of d-electrons (Mn$^{3+}$ → Mn$^{4+}$). **d-e,** Schematic of photoexcited inter-site transitions of the Mn d-electrons in the strain engineered LCMO lattice. The dashed rectangle represents the enhanced orthorhombic unit cell in the a-b plane. The Mn$^{3+}$ e$_g$ electrons can be excited to the empty orbital of Mn$^{4+}$ ions.

**Figure 2 | Photo-induced insulator to metal phase transition in strain-engineered thin film.** With decreasing temperature, photoexcitation (sub-50 fs 1.5 eV pulses at a fluence of 2 mJ/cm$^2$) yields a step-like conductivity progression, until it achieves a conductivity of 1,200 (Ω$^{-1}$cm)$^{-1}$ at 80 K, corresponding to the full "hidden" metallic state. This is more clearly shown in Fig. 1a, with the photoexcitation data plotted as blue dots. Hence, photoexcitation (in zero applied magnetic field) stabilizes the ferromagnetic metallic phase. With increasing temperature (without photoexcitation), the conductivity returns to the insulating phase at 130 K.

**Figure 3 | Pulse-to-pulse conductivity changes and single shot dynamics. a,** Pulse-to-Pulse resistivity change of La$_{2/3}$Ca$_{1/3}$MnO$_3$ film at 80 K, pulse fluence 4 mJ/cm$^2$. **b,** Fluence dependent (1-4 mJ/cm$^2$) pulse-to-pulse conductivity (solid lines) saturates to different final conductivity values, indicative of a cooperative effect dependent on the excitation density. **c,** The saturated photoinduced THz conductivity as a function of the average number of photons absorbed per active Mn$^{3+}$ atom. The colors correspond to the different intensities in **b**. **d,** Ultrafast pulse-to-pulse reflectivity change dynamics (ΔR/R) at 1.7 eV following 1.5 eV excitation. The inset shows the switching dynamics at selected laser shots, which are line-cuts of the 2D plot with the corresponding colors of the dashed lines. **e-j,** The corresponding pulse-to-pulse optical image in the photoinduced metastable metallic state (color scale denotes the normalized photo-induced reflectivity change in visible spectral range). Dashed ellipse in **j** indicates the FWHM of the Gaussian beam.

**Figure 4 | Ginzburg-Landau free-energy phenomenology.** Plot of the free energy with the form $F = aQ^2 + bQ^4 + cM^2 + dM^4 + eM^2Q^2$, where Q is the lattice distortion, M the magnetization, a = a$_0$(T-T$_{CO}$), c = c$_0$(T-T$_m$) with T$_{CO}$ and T$_m$ the charge ordering and ferromagnetic ordering temperatures, respectively. The quantity e is the magnetic-lattice coupling between the order parameters Q and M. **a-b** depicts a contour plot of the free energy with increasing coupling strength at 100K. The red lines (A-B-A') are contours plotted as lineouts (**c-d**) at various temperatures.

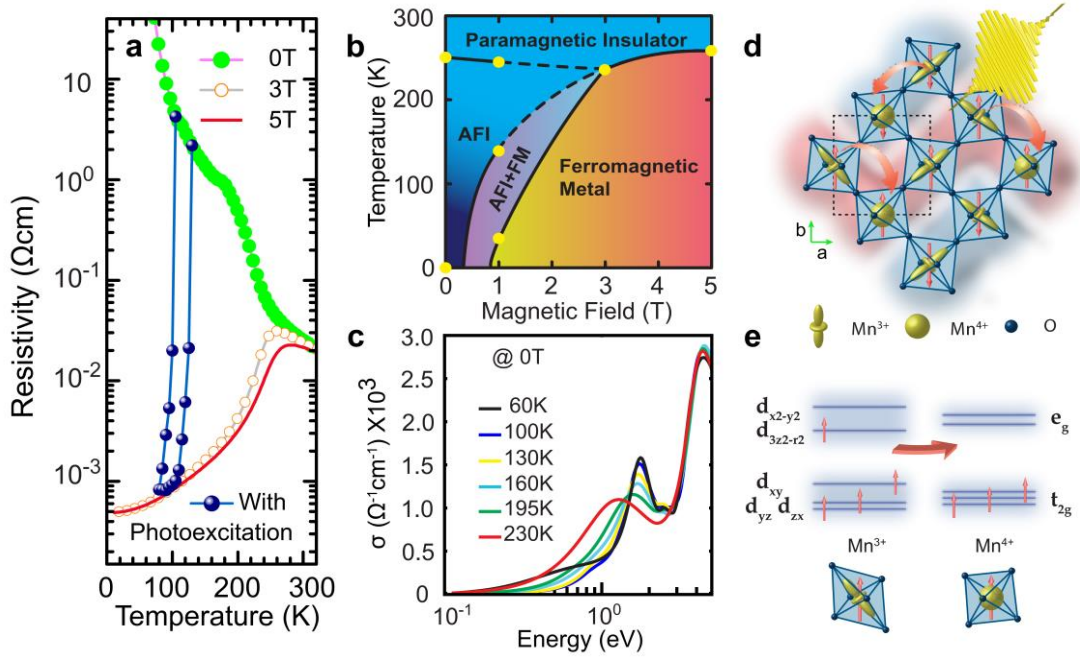

Figure 1

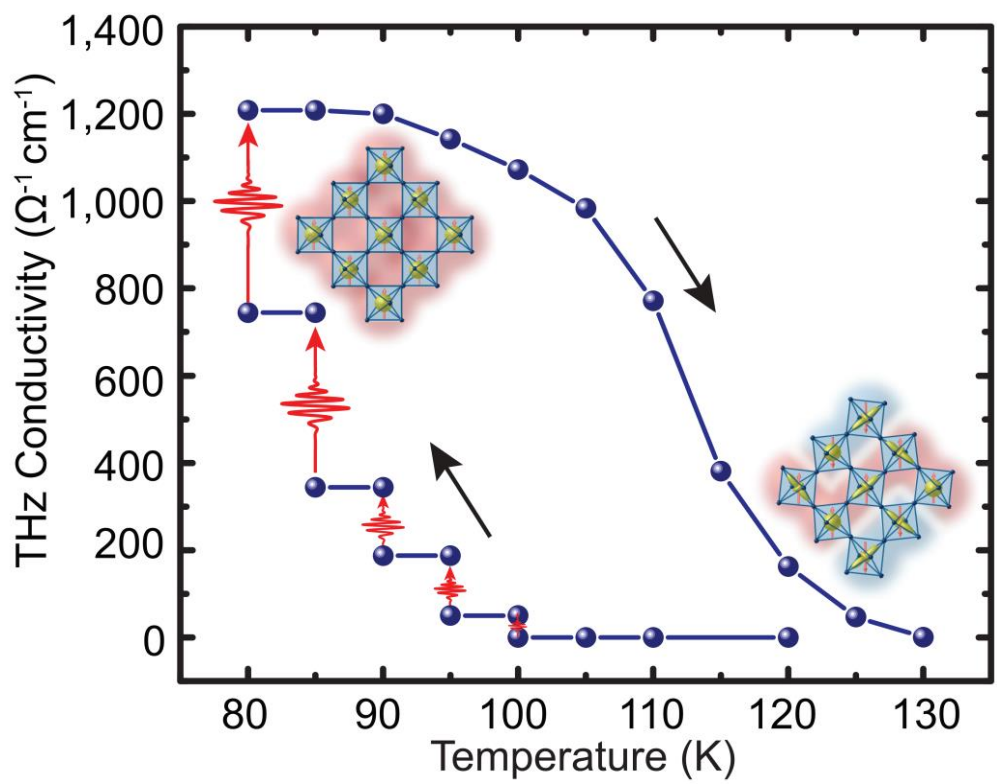

**Figure 2**

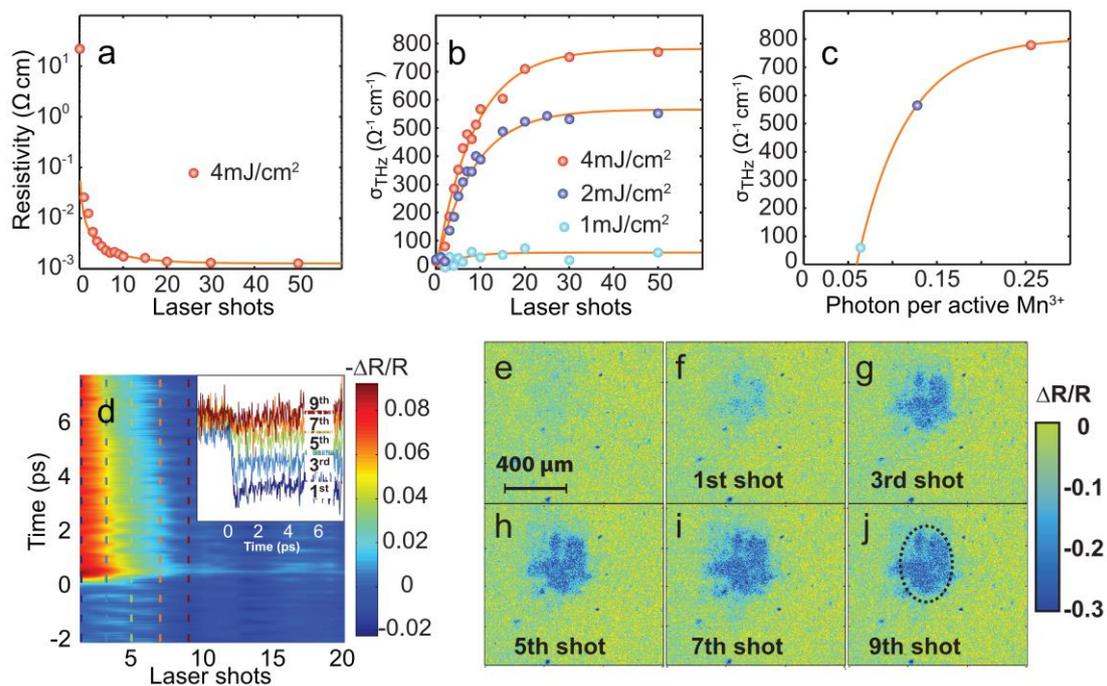

**Figure 3**

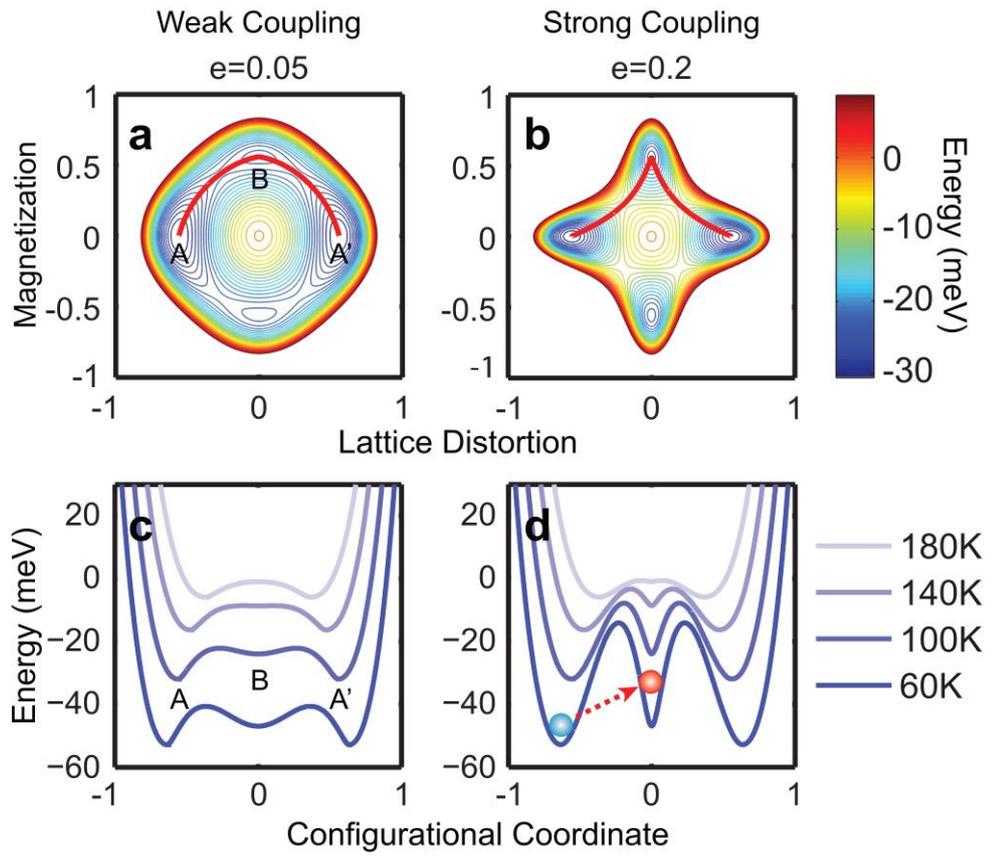

**Figure 4**

**Methods and Supplementary information**

**Sample Growth and Characterization:**

La$_{2/3}$Ca$_{1/3}$MnO$_3$ films with thicknesses of 15 and 30 nm were grown on NdGaO$_3$ (001) substrates by pulsed laser deposition (PLD), using a 248 nm KrF excimer laser with laser energy density of 2 J/cm$^2$ and frequency of 5 Hz. The temperature and oxygen pressure during the deposition were set at 735 °C and 40 Pa, respectively. In order to optimize the oxygen content and to enhance the epitaxial strain coherence, the films were ex-situ annealed at 780 °C in a flowing O$_2$ atmosphere for 6 hours. The structures of the films were analyzed by x-ray diffraction (XRD) using Cu-K$\alpha$ ($\lambda$=1.5406 Å, Panalytical X'Pert) radiation, including high-resolution off-specular x-ray reciprocal space mapping (RSM) measurements at (116) reflection. The surface morphologies of the films were characterized by the atomic force microscopy (AFM, Vecco, MultiMode V). The resistivity measurements were performed on Quantum Design physical property measurement system (PPMS) using standard four-probe technique.

**Supplementary Figure 1 a and b** show AFM images of the 15 nm and 30 nm films, while **c and d** show the XRD and reciprocal space of the 30nm film. The films adopt a step-flow or layer-to-layer growth mode confirmed by the observation of clear atomic steps and terraces in the AFM images. Not only the surface but also the LCMO/NGO interface is smooth, evidenced by the appearance of periodic interference fringes in the XRD profile. The LCMO (116) peak is sharp with the same in-plane Q [110] as that of NGO (116), indicating that the film is of high crystallinity and coherently strained on the substrate.

**Supplementary Figure 2** shows the resistivity of three La$_{2/3}$Ca$_{1/3}$MnO$_3$ films. **(a)** 15 nm film which has not been ex-situ annealed. This film is almost strain free and undergoes an insulator-metal transition at T$_c$=210 K. **(b)** This is the data for the 30 nm annealed film presented in the manuscript. The slight bump around 200 K in 0 T-resistivity may be indicative of the phase boundary between AFI-dominant and FM-dominant states in the phase-separated regime, as determined in Ref. 22. **(c)** A thinner 15 nm annealed film is also insulating in the absence of an applied magnetic field. While this film could be switched with an applied magnetic field, short pulse excitation did not result in a stable metallic phase. This is because the 15 nm annealed film possesses an insulating state at low temperature with strain-induced lattice distortion that is more robust than 30 nm film. This robustness is directly reflected by the difference in resistivity curves between the 15 nm and 30 nm annealed films under a magnetic field of 1 Tesla.

**Optical Conductivity Characterization:**

The optical conductivity $\tilde{\sigma}(\omega) = \frac{i\omega}{4\pi}[1 - \tilde{\varepsilon}(\omega)]$ of the LCMO thin films is obtained from the complex dielectric function $\tilde{\varepsilon}(\omega)$ determined by variable angle spectroscopic ellipsometry (VASE). These measurements were performed using two commercial Woollam ellipsometers (IR-VASE and UV-VASE) equipped with home-built UHV cryogenic chambers, covering a wide photon energy range (0.1 – 5 eV). The mid-IR to near-IR (0.05 eV-0.7 eV) spectra was measured by the IR-VASE system based on a Michelson interferometer (Bruker 66 v/S). The near-IR to UV (0.6-6 eV) ellipsometry was performed via a VASE system based on a grating monochromator (Woollam HS-190). Both systems work under ultra high vacuum (UHV) environment covering from 20 K to 400 K.

The optical conductivity spectra are displayed in **Supplementary Figure 3**. **(a)** The 15 nm (as-grown) film exhibits clear spectral weight transfer from a small polaron to a Drude peak with decreasing temperature, consistent with the expected insulator-metal transition. In contrast, the annealed films **(b)** and **(c)** show spectral weight transfer from the 1.5 eV peak to a well-defined 1.7 eV peak as described in the manuscript and in line with insulating behavior.

**THz Conductivity Measurements:**

The output of 1 kHz, 3 W, 35 fs Ti: sapphire regenerative amplifier (Spectra-Physics Spitfire Pro XP) is used to provide 1.55 eV excitation pulses and generate single-cycle THz pulses by optical rectification in a ZnTe crystal. We use standard electro-optical sampling technique to measure the THz transmission through a bare $NdGaO_3$ reference substrate and the LCMO/NGO sample. By calculating the phase and amplitude information of the THz pulses in the frequency domain, the THz conductivity or photoinduced change in the conductivity can be measured.

The results of the optical-pump THz probe measurements that measure the changes in the THz conductivity are shown in **Supplementary Figure 4** for the same three films as **Supplementary Figure 2** and **3**. The intensity of the optical pump is 2 mJ/cm$^2$. For the 15 nm as-grown film (**Supplementary Figure 4 a**), the THz conductivity dynamics are consistent with previous studies[31]. For initial temperatures starting in the insulating phase (i.e. > 240K), the optical excitation induces a rapid 2 ps transient increase in conductivity. It quickly decays (1 ps) into a plateau with much longer lifetime. In the metallic phase, a fast decrease in conductivity from electron-photon equilibration occurs. This is followed by a slow component, corresponding to spin-lattice relaxation[31]. In the 30 nm annealed LCMO film (**Supplementary Figure 4 b**) (in which the photo-induced insulator-to-metal phase transition is demonstrated), there is an evolution of the ultrafast dynamics that resembles the bulk-like 15 nm as-grown LCMO film though the magnitude is smaller. In the fully strained film (15 nm annealed –**Supplementary Figure 4 c**), there is a similar tendency of the sample to transit into the metallic phase. However, it is much weaker, and without persistent photoinduced persistent phase transition. The absence of the photo-induced phase transition in this film is due to the large strength of the elastic strain, such that a persistent metallic phase is no longer energetically favored.

**Single-shot THz Conductivity measurement:**

For the single shot THz conductivity measurements (**Fig. 3 a–c**), the intracavity Pockels cell of the Ti:sapphire laser is disabled to stop the output of continuous pulses at 1 kHz.

To emit a single pulse a trigger signal is sent to enable the Pockels cell. In this way, a single laser pulse perturbs the sample. The subsequent THz conductivity measurement is performed after switching back to the 1 kHz output mode. This process is repeated on a shot-by-shot basis.

**Single-shot Optical Pump-probe Spectroscopy:**

Single-shot pump-probe spectroscopy (**Supplementary Figure 5**) was carried out using a dual-echelon single shot instrument[28, 32]. A single pump pulse excites the sample, and an array of probe pulses, generated by a pair of crossed echelons, probe the center of the pump spot, reflect off the sample, and are imaged on a CCD camera. This generates 400 probe pulses, spaced 23 fs apart. For each pump shot, two images of the echelons are recorded, a signal and background. The ratio of signal and background intensities is related to the pump-probe trace by

$$\frac{\Delta R}{R} = \frac{signal}{background} - 1 \qquad \textbf{Equation 1}$$

The laser system is a 1 kHz laser system, downcounted using a pockels cell/polarizer to 10 Hz, and further gated with shutters. The sample is pumped with a 70 fs, 800 nm pulse, and probed with the output of a non-collinear optical parametric amplifier (NOPA) tuned to 730 nm (1.7 eV) with a 50 fs pulsewidth.

Single-shot optical image (**Supplementary Figure 6**) was taken simultaneously by imaging the white light source irradiated sample onto a CCD camera operating in the visible range.

**Photo-induced Metallic Phase Volume Fraction (Effective Medium Theory):**

The volume fraction dynamics as a function of laser shots may be described by the following equation:

$$df_m = \frac{1}{\alpha} \cdot (1 - f_m) \cdot dN \qquad \textbf{Equation 2}$$

yielding,

$$f_m(N) = 1 - (1 - f_1) \cdot \exp\left(-\frac{N-1}{\alpha}\right) \qquad \textbf{Equation 3}$$

where $f_m$ is the volume fraction of the metallic phase, $1/\alpha$ is the increase rate of photo-induced conductivity from laser shot to shot, N is the total number of laser pulses hitting the sample, and $f_1$ is the volume fraction of metallic phase after the first laser shot.

We use the Effective Medium Theory (EMT)[33] to describe the conductivity of the persistent photo-induced state as follows:

$$f_m \frac{\sigma_m - \sigma_{eff}}{\sigma_m + (d-1)\sigma_{eff}} + (1 - f_m)\frac{\sigma_i - \sigma_{eff}}{\sigma_i + (d-1)\sigma_{eff}} = 0 \qquad \textbf{Equation 4}$$

where $\sigma_m$ is the full metallic state conductivity, $\sigma_i$ is the conductivity in the insulating phase, $\sigma_{eff}$ is the effective conductivity, and d is the dimension of sample. Since the thickness of the LCMO film is 30 nm, we take two-dimensional form of the equation (d=2). This yields (taking $\sigma_i = 0$, and $f_1$ to be 0.51 – i.e. percolation is reached after the first shot[34] (see **Fig. 3a**)),

$$\sigma_{eff} = \sigma_m \cdot (2f_m - 1) \qquad \textbf{Equation 5}$$

Plugging equation (2) into equation (4) yields:

$$\sigma(N) = \sigma_m \left[ 1 - 2(1 - f_1) \cdot \exp\left(-\frac{N-1}{\alpha}\right) \right] \qquad \textbf{Equation 6}$$

Equation 5 was used to fit the data in **Fig. 3b** of the main text. For all three fluences α has an intensity independent value of ~8.5. From shot to shot, the change in the conductivity corresponds to an increase in the metallic volume fraction with a corresponding decrease in the insulating volume fraction.

**Ginzburg-Landau model with biquadratic order parameter coupling:**

We use phenomenological Ginzburg-Landau theory to explain the persistent photoinduced insulator-metal transition in strained LCMO through magnetic-lattice (ML) coupling. The existence of such ML coupling makes it feasible to trigger phase transition by stimuli such as magnetic field, pressure, strain and photoexcitation, in particular. The Ginzburg-Landau free energy includes elastic, magnetic energy and the ML coupling term. The order parameters of the LCMO film are the lattice distortion Q (Jahn-Teller distortion, or tilting and rotation mode of the oxygen octahedron) and magnetization M. The full expression of the Ginzburg-Landau free energy may be described in a 2D space of Q and M by the equation below.

$$F = aQ^2 + bQ^4 + cM^2 + dM^4 + eM^2Q^2 \quad \begin{aligned} a &= a_0(T - T_{CO}) \\ c &= c_0(T - T_m) \\ a_0, b_0, c, d &> 0 \end{aligned} \quad \textbf{Equation 7}$$

where $T_{CO}$ is the charge ordering temperature 240 K, $T_m$ is the onset temperature of the ferromagnetism without the charge ordered phase (lattice strain) 200 K. $T_m$ strongly depends on the applied the H field on the sample. The value of e denotes the ML coupling of the order parameters Q and M. At lower temperature (T<105 K), the ML coupling should have a pronounced influence on the free energy landscape, corresponding to a "locking" mechanism from the coupling of the spin and lattice degrees of freedom. Within this temperature range, the photo-induced phase transition can proceed and have persistent nature. To capture the main feature of the balance of the competing lattice and spin degrees of freedom, we assign values to parameters in the Landau-Ginzburg free energy, $a_0$=0.1826 meV/K, b=0.0256 meV, $c_0$=0.1917 meV/K, d=0.0192 meV, e∈[0.05, 0.20] meV. These values are scaled estimates of the energy from calculations of the activation energy. The activation energy is determined by the laser excitation flux, and the temperature (130~140 K) resulting in a return to the pristine AFI state through thermal fluctuations. **Supplementary Figure 7 a-d** gives the free energy contours with various values of ML coupling constant at different temperatures. Each energy contour has four minima (two global minima A and A', and two local minima B and B'). In the case of the strained LCMO film, the global minimum A (A') corresponds to the antiferromagnetic insulating phase, where Q≠0 and M=0, representing the ground state of the system. B (B') is the metastable ferromagnetic metallic phase, which competes with the insulating phase. **Supplementary Figure 7 e.1-e.4** shows the energy landscape as a function of configurational coordinate (lattice distortion) along the highlighted (red) path (A-B-A') in the contour map, with changing temperature and ML coupling constant (e=0.05, 0.07, 0.13, 0.20). As the ML coupling constant increases or the temperature decreases, the energy barrier between global minimum A (A') and local minimum B (B') increases. When the coupling e is small (or temperature is high), the energy barrier between A (A') and B (B') is small. The ferromagnetic metallic phase B (B') is unstable because the shallow minimum. Even though B (B') is a local minimum, with energy higher than true ground state, the energy barrier is such that a metastable state is formed. Photoexcitation provides a path to reach the metastable metallic state. With increasing temperature and thermal fluctuations, the LCMO film will be able to return to insulating phase.

31. Averitt, R. D. *et al.* Ultrafast Conductivity Dynamics in Colossal Magnetoresistance Manganites. *Physical Review Letters* **87,** 017401 (2001).

32. Poulin, P. R. & Nelson, K. a. Irreversible organic crystalline chemistry monitored in real time. *Science (New York, N.Y.)* **313,** 1756–60 (2006).

33. Choy, T. C. *Effective medium theory: principles and applications*. (Oxford University Press, 1999).

34. Hilton, D. *et al.* Enhanced Photosusceptibility near Tc for the Light-Induced Insulator-to-Metal Phase Transition in Vanadium Dioxide. *Physical Review Letters* **99,** 226401 (2007).

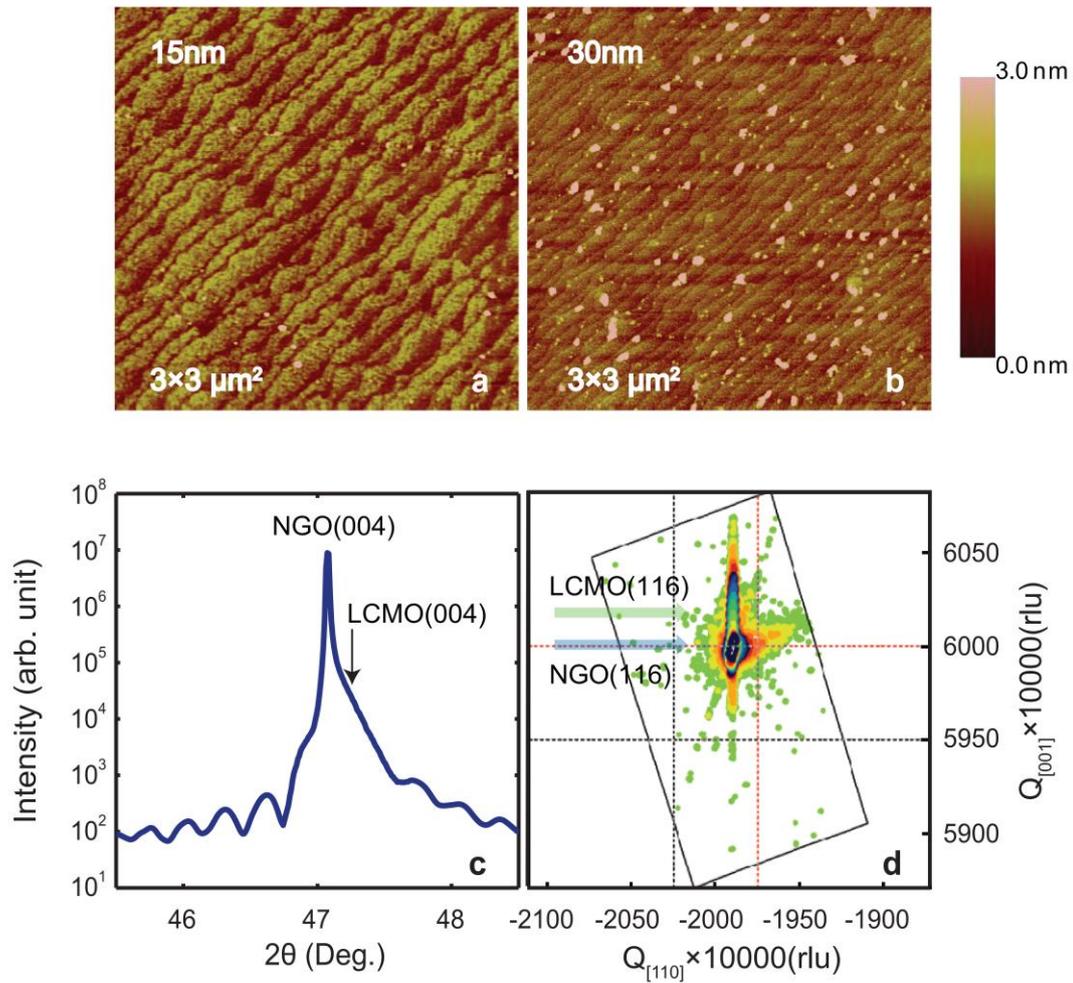

**Supplementary Figure 1 | La$_{2/3}$Ca$_{1/3}$MnO$_3$ thin film characterization. a-b,** AFM images of the annealed LCMO/NGO (001) films with thickness of 15 nm and 30 nm. **c,** X-ray diffraction (XRD) linear scan profile near the (004) peak of LCMO film and NGO substrate. **d,** RSM (reciprocal space mapping) near the (116) peak of the annealed 30 nm LCMO/NGO (001) film.

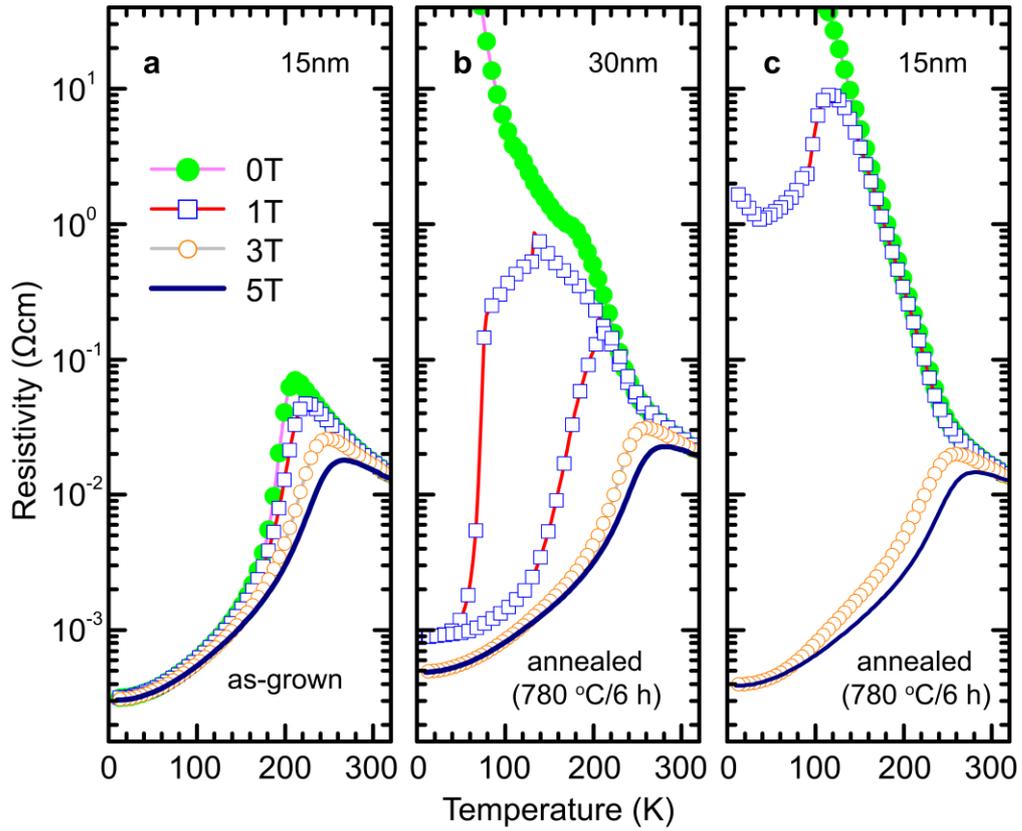

**Supplementary Figure 2 | Resistivity-Temperature curves of $La_{2/3}Ca_{1/3}MnO_3$ thin films of different anisotropic strain level. a,** 15 nm as-grown (no strain) $LaCa_{1/3}MnO_3$ film. **b,** 30 nm (intermediate strain-level) and **c,** 15 nm (high strain) strain-engineered $La_{2/3}Ca_{1/3}MnO_3$ FILM, under various magnetic field (0 T ~ 5 T). The insulating state is achieved by post-annealing the film at 780 °C for 6 hours.

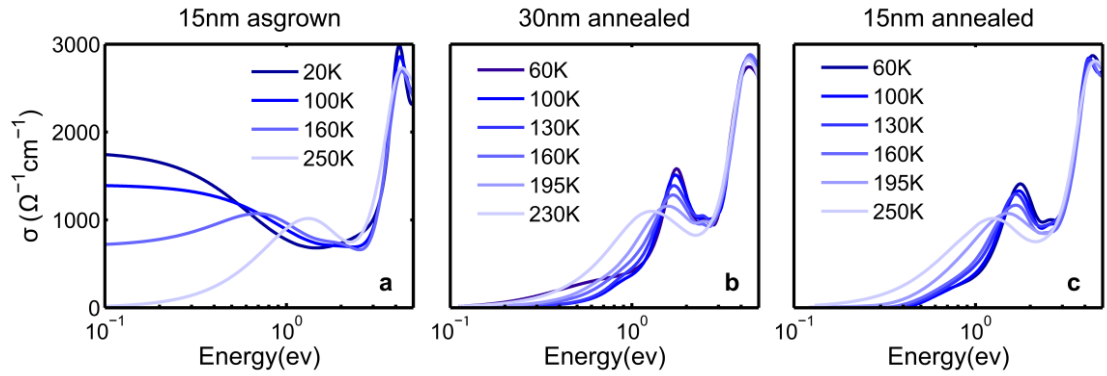

**Supplementary Figure 3 | Temperature dependent optical conductivity ($\sigma_1$) of La$_{2/3}$Ca$_{1/3}$MnO$_3$ thin films. a,** 15 nm as-grown sample shows spectral weight transfer from a small polaron peak to a Drude response, corresponding to the thermal-driven semiconductor to ferromagnetic metal transition. **b and c,** in the strain-engineered thin film, the small polaron peak at high temperature evolves into sharp d-d charge transfer peak at T < 200 K, coincident with the charge ordering temperature. It corresponds to a transition from semiconductor to antiferromagnetic insulator.

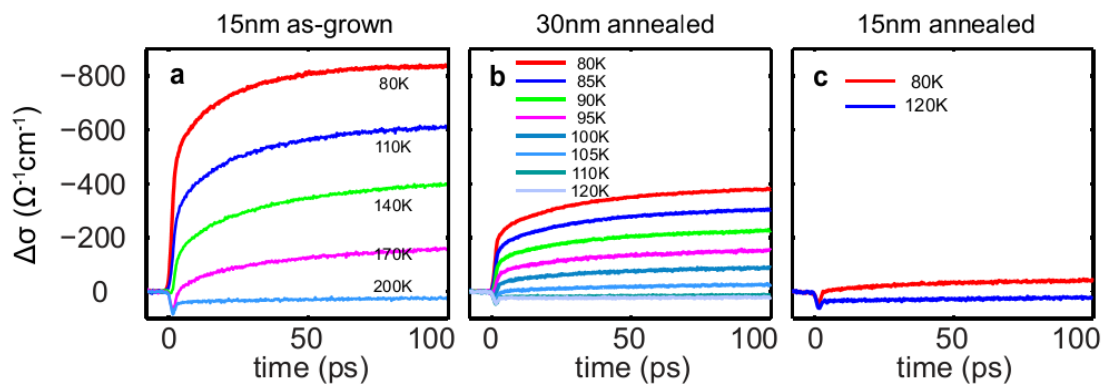

**Supplementary Figure 4 | Ultrafast optical pump THz probe dynamics of $La_{1/3}Ca_{2/3}MO_3$ thin films.** Conductivity dynamics as a function of time following photoexcitation with 35 fs laser pulses ($\lambda$ = 800 nm, fluence 2 mJ/ cm$^2$). **a,** At T < 200K, 15 nm as-grown LCMO film shows conductivity dynamics with opposite sign to T > 200 K dynamics and a slow component, corresponding to spin-lattice relaxation. **b,** THz conductivity dynamics of the photo-switched 30 nm annealed LCMO film at different temperatures. The saturated metallic state shown in **Fig. 2**, shows the same THz conductivity dynamics as the 30 nm as-grown film in metallic state, suggesting the photo-induced metallic state in 30nm annealed sample is ferromagnetic. **c,** 15 nm annealed LCMO film shows conductivity dynamics that corresponds to an insulating phase.

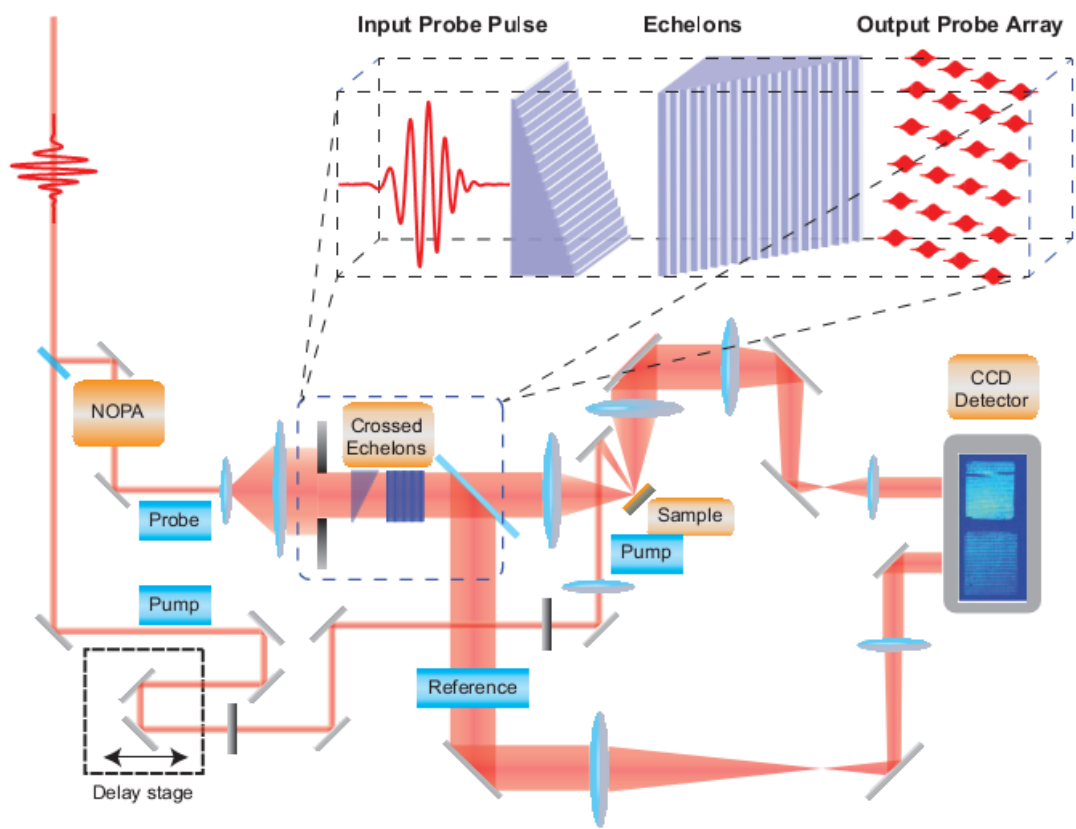

**Supplementary Figure 5 | Schematic of single-shot pump-probe spectroscopy:** Schematic of the dual-echelon single-shot pump-probe setup. A pump pulse (800 nm, 70 fs) is incident on the sample. The probe is expanded 10x, collimated, and pass through crossed thin and thick echelons with 20 steps each, (step thickness of 15 and 300 micron steps, respectively), generating a 20 by 20 array of pulses spaced 23 fs apart. The probe array is focused onto the sample and imaged onto a CCD camera (labeled signal). A beam splitter and separate imaging arm (labeled reference) bypasses the sample and images the grid directly on the camera to account for intensity fluctuations in the probe array.

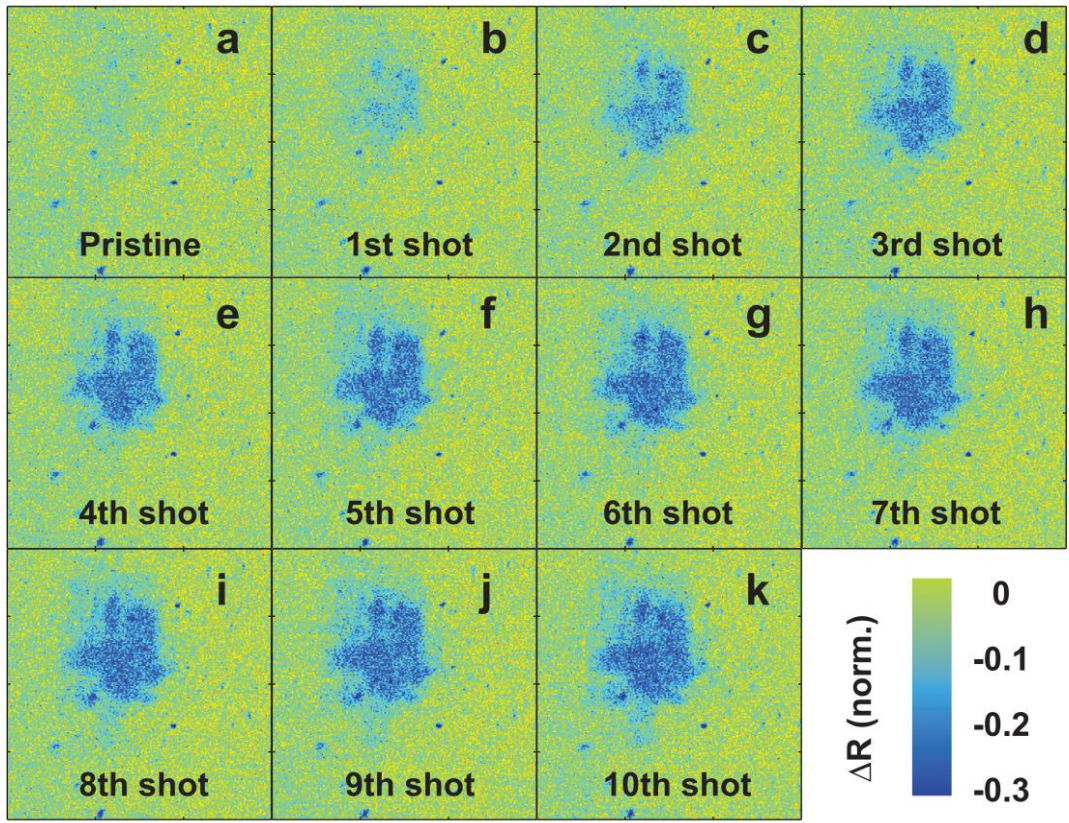

**Supplementary Figure 6 | Single-shot optical image. a-k**, a complete set of the pulse-to-pulse optical image of the photo-induced metastable metallic state (color scale denotes the normalized photo-induced reflectivity change in visible spectral range).

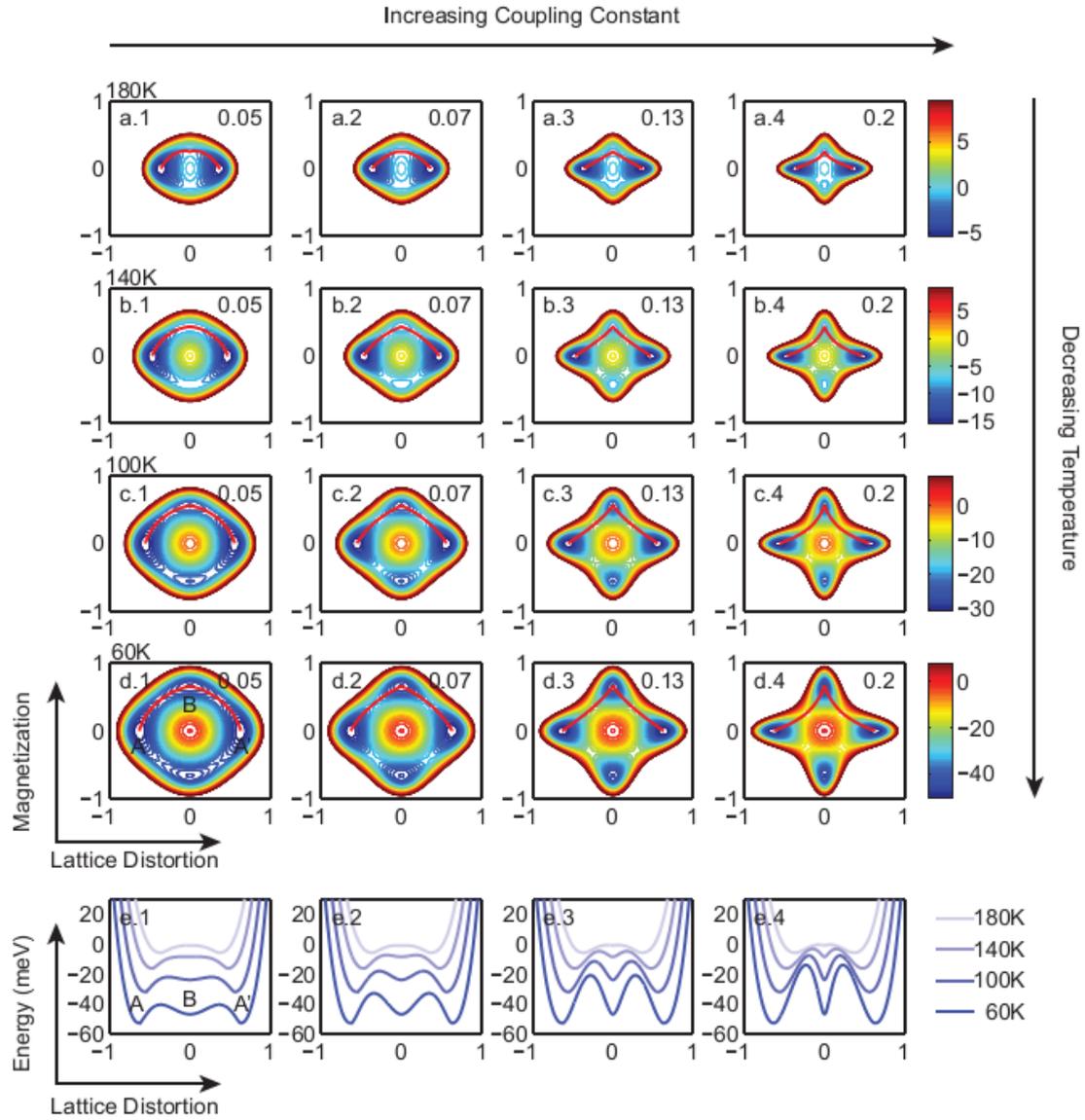

**Supplementary Figure 7 | Ginzburg-Landau free-energy phenomenology. a-d,** Contour plots of the free energy with increasing magnetic-lattice coupling coefficient (left to right) and temperature (top to bottom), described by Equation 7 in Methods. Each contour plot has two global minima (A and A') and two local minima (B), which are separated by an energy barrier determined by the magnetic-lattice coupling strength. The red lines (A-B-A') are contours plotted as lineouts in **(e1-e4)**.

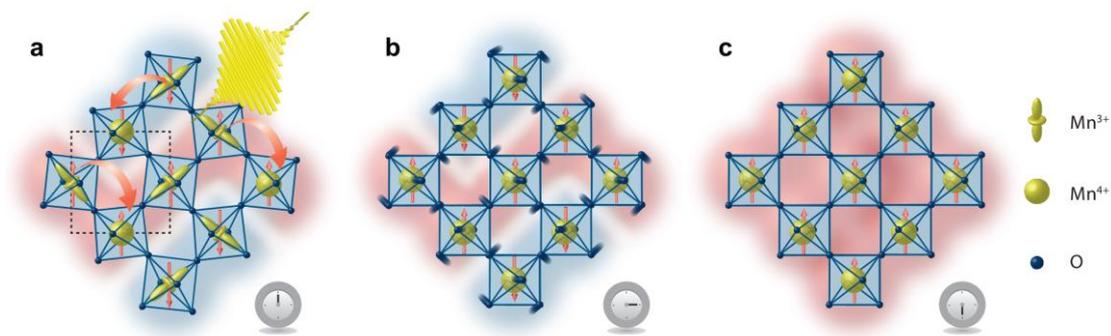

**Supplementary Figure 8 | Detailed schematic of strained LCMO response upon 1.55 eV short laser pulse excitation. a,** 1.55 eV photo-excitation delocalizes d-electrons associated with the $Mn^{3+}$ sites. Delocalized d-electrons hop along the zig-zag path with identical core spin orientation (red shading, spin up; blue shading, spin down). **b,** Delocalization of the d-electron relaxes Jahn-Teller distortions and/or octahedral rotations. This reduces the orthorhombicity induced by elastic strain that favors the charge-ordered antiferromagnetic phase as the ground state. **c,** Transient delocalization of electrons and relaxation of lattice distortion affects the magnetic exchange interaction between neighboring zig-zag chains (blue and red paths) resulting in the stable photo-induced ferromagnetic metallic phase. Recovery to the insulating phase at low temperatures is forbidden due to the strong magnetic-lattice interaction as described by Ginzburg-Landau phenomenology.